\documentclass[12pt]{iopart}
\usepackage{amssymb}
\usepackage{graphicx}

\begin{document}
\newcommand{\anfEngl}[1]{``#1''} 
\newcommand{\ens}[0]{\ensuremath} 
\newcommand{\GXOR}[0]{\mathrm{GXOR}}
\newcommand{\GBXOR}[0]{\mathrm{GBXOR}}
\newcommand{\AsymCSS}[0]{\mathrm{AsymCSS}}
\newcommand{\aeg}[0]{\stackrel{a.e.}{=}}
\newcommand{\ket}[1]{\ensuremath{|#1 \rangle}} 
\newcommand{\bra}[1]{\ensuremath{\langle #1|}} 
\newcommand{\pr}[1]{\ens{\ket{#1}\bra{#1}}} 
\newcommand{\x}[0]{\ens{\otimes}} 
\newcommand{\impl}[0]{\ens{\Rightarrow}} 
\newcommand{\eqv}[0]{\ens{\Leftrightarrow}} 
\newcommand{\nach}[0]{\ens{\rightarrow}} 
\newcommand{\nGr}[0]{\ens{{n \rightarrow \infty}}} 
\newcommand{\Proof}[0]{\emph{Proof: }} 
\newcommand{\Mge}[2]{\ens{\left\lbrace #1|\,#2 \right\rbrace}}
\newcommand{\Mg}[1]{\ens{\left\lbrace #1 \right\rbrace}}
\newcommand{\Mgn}[2]{\ens{\Mg{#1,\dots,#2}}} 
\newcommand{\MgN}[1]{\ens{\Mg{0,\dots,#1}}} 
\newcommand{\MgE}[1]{\ens{\Mg{1,\dots,#1}}} 
\newcommand{\norm}[1]{\ens{\left\|#1\right\|}} 
\newcommand{\betrag}[1]{\ens{\left|#1\right|}} 
\newcommand{\Folge}[3]{\ens{(#1_#2)_{#2 \in #3}}}
\newcommand{\BE}[0]{\hfill $\Box$} 
\newcommand{\Fkt}[3]{\ens{#1 : #2 \nach #3}} 
\def\Eins{{\leavevmode{\rm 1\ifmmode\mkern -4.4mu\else\kern -.3em\fi I}}}
\newcommand{\eps}[0]{\ens{\varepsilon}}
\newcommand{\ksi}[0]{\ens{\xi}}
\newcommand{\cF}[0]{\ens{\mathcal{F}}}
\newcommand{\cH}[0]{\ens{\mathcal{H}}}
\newcommand{\cS}[0]{\ens{\mathcal{S}}}
\newcommand{\cW}[0]{\ens{\mathcal{W}}}
\newcommand{\N}[0]{\ens{\mathbb{N}}}
\newcommand{\R}[0]{\ens{\mathbb{R}}}
\newcommand{\C}[0]{\ens{\mathbb{C}}}
\newcommand{\Z}[0]{\ens{\mathbb{Z}}}
\newcommand{\Sbd}[0]{\ens{\cS_{\mathrm{bd}}}}
\newcommand{\Sbdd}[0]{\ens{\Sbd^{(d)}}}
\newcommand{\lm}[2]{\ens{\left(#1,#2\right)}}
\newcommand{\Bn}[0]{\ens{B_n}}
\newcommand{\Bnd}[0]{\ens{B_n^{(d)}}}
\newcommand{\Snd}[0]{\ens{S_n^{(d)}}}
\newcommand{\iE}[0]{\ens{\mathrm{i}}} 
\newcommand{\Alm}[0]{\ens{(A_{lm})_{l,m = 0}^{d-1}}}
\newcommand{\Al}[0]{\ens{(A_{l*})_{l = 0}^{d-1}}}
\newcommand{\Am}[0]{\ens{(A_{*m})_{m = 0}^{d-1}}}
\newcommand{\Aplm}[0]{\ens{(A^\prime_{lm})_{l,m = 0}^{d-1}}}

\newtheorem{Definition}{Definition}
\newtheorem{Theorem}{Theorem}
\newtheorem{Lemma}{Lemma}

\title[Asymptotic correctability of Bell-diagonal qudit states]{Asymptotic correctability of Bell-diagonal
qudit states and lower bounds on tolerable error probabilities in quantum cryptography}
\author{Kedar S Ranade and Gernot Alber}\ead{Kedar.Ranade@physik.tu-darmstadt.de}
\address{Institut f\"ur Angewandte Physik, Technische Universit\"at Darmstadt,\\
  64289 Darmstadt, Deutschland (Germany)}
\date{September 26, 2006}

\begin{abstract}
The concept of asymptotic correctability of Bell-diagonal quantum states is generalised to elementary
quantum systems of higher dimensions. Based on these results basic properties of quantum state
purification protocols are investigated which are capable of purifying tensor products of Bell-diagonal
states and which are based on $B$-steps of the Gottesman-Lo-type with the subsequent application of a
Calderbank-Shor-Steane quantum code. Consequences for maximum tolerable error rates of quantum
cryptographic protocols are discussed.
\end{abstract}


\section{Introduction}
Quantum state purification protocols which are based on local operations and classical communication and
which are capable of purifying tensor products of Bell-diagonal quantum states are of considerable current
interest in the area of quantum cryptography. This may be traced back to the fact that the security
analysis and questions concerning achievable secret-key rates of many quantum cryptographic protocols
are based on basic properties of such quantum-state purification protocols \cite{SP,GL}.
So far, a satisfactory understanding of such protocols has already been obtained in qubit-based scenarios. 
In particular, it was demonstrated that powerful quantum-state purification protocols can be developed
for tensor products of Bell-diagonal states by combining a sufficiently large number of purification
steps involving classical two-way communication, so called $B$-steps \cite{GL}, with subsequent
quantum error correction based on Calderbank-Shor-Steane (CSS) codes \cite{CSS} which involve
classical one-way communication only. Furthermore, the asymptotic properties of these protocols for
large numbers of $B$-steps can be analysed in a convenient way by characteristic exponents which govern
the relation between bit- and phase errors \cite{KSR}. Based on such an analysis it is straightforward,
for example, to determine maximally tolerable bit-error probabilities of quantum cryptographic protocols of
the prepare-and-measure type whose security analysis can be reduced to the purification of Bell-diagonal
qubit states \cite{GL,KSR,Acin,Ch02}.
Contrary to qubit-based scenarios, elementary properties of quantum-state purification protocols are
still rather unexplored in quantum cryptographic contexts in which the transfer of quantum information
is based on higher-dimensional elementary quantum systems, so called qudits.

Recently, some qudit-based quantum cryptographic protocols were developed whose security analysis
can be related to basic properties of quantum-state purification protocols capable of purifying tensor
products of generalized Bell-diagonal quantum states \cite{Ch05,NA,NRA}. Motivated by these current
developments in this paper the asymptotic properties of  qudit-based quantum-state purification
protocols are investigated which involve $B$-steps and the subsequent application of a CSS code fulfilling
the Shannon bound of Hamada \cite{Ham}.
For this purpose, the previously developed concept of asymptotic correctability is generalized to
arbitrary-dimensional elementary quantum systems and corresponding relevant exponents are determined
which govern the relation between dit-and phase errors for large numbers of purification steps (compare
with theorem \ref{AsymCorr}). In quantum cryptographic applications the phase-error probabilities are not
accessible to direct measurement, but they have to be estimated on the basis of the measured qudit-error
probabilities. For this purpose it is convenient to start a purification protocol with a local unitary mixing
transformation which homogenises the phase errors associated with each possible dit error.
The asymptotic correctability under the resulting quantum-state purification protocol can be determined
in a rather straightforward way (compare with theorem \ref{BndCorr}). This latter result is particularly
useful for determining lower bounds on maximally tolerable qudit-error probabilities of quantum
cryptographic protocols whose security analysis can be reduced to the asymptotic correctability under
these latter quantum-state purification protocols. 

This paper is organized as follows: In section \ref{AllgDef} basic notions of qudit-systems, such as
the definition of generalised Bell states, are summarized. Section \ref{SecAsymCorr} is devoted to the
definition of asymptotic correctability of general quantum state purification protocols which involve tensor
products of generalized Bell-diagonal qudit states. In particular, theorem \ref{AsymCorr} relates this
asymptotic correctability to basic properties of exponents which govern the relation between dit and
phase errors. Section \ref{SecPurProt} specializes these results to purification protocols which start
with a local mixing operation followed by generalised $B$-steps and a subsequent application of a CSS
quantum code. In Section \ref{SecQKD} lower bounds on maximally tolerable dit-error probabilities of
quantum cryptographic protocols are discussed whose postprocessing can be reduced to the analysis of
such purification protocols.

\section{Quantum systems of dimension $d$}\label{AllgDef}
We consider a quantum system of dimension $d$, which is called a qudit.
A certain orthonormal basis of the associated Hilbert space $\cH = \C^d$ is labelled
by the elements of the set $\Z_d := \MgN{d-1}$, which are representatives
of the ring of residue classes $\Z/d\Z$, i.\,e. we consider all operations modulo $d$;
we denote addition and subtraction by \anfEngl{$\oplus$} and \anfEngl{$\ominus$}, respectively.
We further denote $\Z_d^* := \Z_d \setminus \Mg{0}$.\footnote{Unless $d$ is a prime, $\Z_d^*$ does not
represent the set of invertible elements of $\Z/d\Z$.} In analogy to the abbreviation \anfEngl{bit} for
\anfEngl{binary digit} we use the term \anfEngl{dit} for \anfEngl{$d$-ary digit}.
\par We will need the notion of a probability distribution on $d$ elements, which can be
identified with normalised $d$-tuples of non-negative real numbers. For convenience, we denote
the set of such tuples by
\begin{equation}
  \cW_d := \Mg{(p_0,\dots,p_{d-1}) \in \R^d \,\left|\,\sum_{i = 0}^{d-1} p_i = 1;\,
    p_i \geq 0 \quad\mathrm{for\,all\,\,} i \right.}.
\end{equation}
For such a probability distribution $p = (p_0,\dots,p_{d-1}) \in \cW_d$ the Shannon entropy is defined by
\begin{equation}
  H_d(p) := - \sum_{i = 0}^{d-1} p_i \log_d p_i = - (\ln d)^{-1} \sum_{i = 0}^{d-1} p_i \ln p_i.
\end{equation}
The Hilbert space of a pair of qudits, i.\,e. $\cH \x \cH$, has a basis of maximally entangled
states, which we call the \emph{(generalised) Bell basis} of this system. It is defined by \cite{ADGJ}
\begin{equation}
  \ket{\Psi_{lm}} := \frac{1}{\sqrt{d}}\left[\sum_{k = 0}^{d-1} z^{lk} \ket{k} \ket{k \ominus m}\right]
  \mathrm{\quad for \quad} l,\,m \in \Z_d,
\end{equation}
where $z := \exp(2\pi\iE/d)$ is the principal root of unity of order $d$. We denote the associated
density matrices by $(l,m) := \pr{\Psi_{lm}}$. We will frequently use classical mixtures of generalised
Bell states, i.\,e. states of the form
\begin{equation}
  \rho = \sum_{l,m=0}^{d-1} A_{lm} \pr{\Psi_{lm}}, \quad\mathrm{where}\quad \Alm \in \cW_{d \times d}.
\end{equation}
Such mixtures we will identify with their coefficient matrix\footnote{The coefficient matrix is not
a density matrix on a Hilbert space.}, so that we can write
\begin{equation}
  \rho = \Alm = \left(\begin{array}{cccc}
    A_{00} & A_{01} & \dots & A_{0,d-1} \\
    A_{10} & A_{11} & \dots & A_{1,d-1} \\
    \vdots & \vdots & \ddots & \vdots   \\
    A_{d-1,0} & A_{d-1,1} & \dots & A_{d-1,d-1}\\
  \end{array}\right).
\end{equation}
The only condition on the entries is, that they form a probability distribution on $\Z_d \times \Z_d$,
i.\,e. that all $A_{lm}$ are non-negative and sum up to one. The set of all such mixtures of
generalised Bell states will be denoted by \Sbdd.
\par We will consider $\ket{\Psi_{00}}$ as the reference state for purification, so that we can interpret
$l$ and $m$ as phase and dit errors, respectively. The columns of the coefficient matrix thus represent
different dit values, whereas the rows represent different phase values. Marginal distributions of dit
and phase errors are therefore given by
\begin{equation}
  A_{*m} := \sum_{l=0}^{d-1} A_{lm} \quad\mathrm{for}\quad m \in \Z_d 
 \qquad\mathrm{and}\quad A_{l*} := \sum_{l=0}^{d-1} A_{lm} \quad\mathrm{for}\quad l \in \Z_d.
\end{equation}
A generalised XOR operation on two qudits, the source and the target, is defined by
$\GXOR \ket{k} \ket{l} := \ket{k} \ket{k \ominus l}$ \cite{ADGJ}. The bilateral version applied
to two pure generalised Bell states $(l_1,m_1)$ and $(l_2,m_2)$ yields
\begin{equation}
   \GBXOR\bigl[(l_1,m_1) \x (l_2,m_2)\bigr] = (l_1 \oplus l_2, m_1) \x (l_2, m_1 \ominus m_2).
\end{equation}
Another mathematical tool which we use is the so-called $p$-norm for tuples of fixed length, where
$p \in [1;\infty]$. For $x = (x_0, x_1, \dots, x_{d-1}) \in \C^d$ it is defined by
\begin{equation}
  \norm{x}_p := \left(\sum_{i = 0}^{d-1} \betrag{x_i}^p \right)^{1/p}
\end{equation}
for $p \in [1;\infty)$ and $\norm{x}_\infty := \max\Mge{\betrag{x_i}}{i \in \Z_d}$. We have
$\norm{x}_p \geq \norm{x}_q$ for $p \leq q$ and $\lim_{p \rightarrow \infty} \norm{x}_p = \norm{x}_\infty$.
If $\betrag{x_i} \leq 1$ for all $i$ (which e.\,g. is the case, if $x \in \cW_d$), also
$\norm{x}_p^p \geq \norm{x}_q^q$ holds. Of particular interest is the fact that the $2$-norm is
invariant with respect to a discrete Fourier transform.

\section{Asymptotic correctability for qudit systems}\label{SecAsymCorr}
In this section we consider entanglement purification protocols and their properties. We assume that
two distant parties, Alice and Bob, share a large amount of mixtures of generalised Bell states, i.\,e.
their joint state is $\rho^{\x n}$ for $\rho \in \Sbdd$ and some large $n \in \N$. They perform two-way
entanglement purification until the use of a CSS code fulfilling the quantum Shannon bound allows them
to extract some pure generalised Bell state, e.\,g. $\ket{\Psi_{00}}$. The quantum Shannon bound is
given by the following theorem.
\begin{Theorem}[Quantum Shannon Bound]\label{QSS}\hfill\\
  Let $d$ be a prime number and consider a state $\rho = \Alm \in \Sbdd$. If
  \begin{equation*}
    \AsymCSS_d\bigl[\Alm\bigr] := 1 - H_d\left[\Am\right] - H_d\left[\Al\right] > 0,
  \end{equation*}
  there exists a CSS code which can correct a tensor product state $\rho^{\x n}$.
\end{Theorem}
\Proof This is an obvious consequence of a theorem by Hamada (\cite{Ham}, Theorem 2). \BE\vspace{0.5cm}\\
Using this bound, we can now define the notion of asymptotic correctability; due to the use of that
theorem, in the following we consider $d$ always to be prime. For $d = 2$, this definition reduces to
the one given in \cite{KSR}.
\par A correction step $\Snd$ of a quantum state purification protocol takes as input a state of the form
$\rho^{\x n}$ and outputs a state of the form ${\rho^\prime}^{\x n^\prime}$, where
$\rho,\,\rho^\prime \in \Sbdd$. In general, $n^\prime \leq n$ and $\rho^\prime$ is supposed to be
more entangled than $\rho$. Occasionally, a step may fail and does not output anything. As we do not
consider distillation rates we can drop the labels $n$ and $n^\prime$. A correction step
will thus be treated as a function on $\Sbdd$, mapping $\Alm$ to $\Aplm$.
\begin{Definition}[Asymptotic correctability]\hfill\\
  Let $\rho = \Alm \in \Sbdd$ and $(\Snd)_{n \in \N}$ be a sequence of possible correction steps
  in an entanglement purification protocol. The state $\rho$ is called \emph{asymptotically \Snd-correctable},
  if the inequality $\AsymCSS\bigl[\Snd(\rho)\bigr]~>~0$ holds for all $n \geq N_0$, where $N_0 \in \N$.
  We call $\rho$ \emph{asymptotically non-correctable} under the sequence
  $(\Snd)_{n \in \N}$, if $\AsymCSS\bigl[\Snd(\rho)\bigr]~\leq~0$ holds for $n \geq N_0$ for some $N_0 \in \N$.
\end{Definition}
We now want to generalise the criterion for asymptotic correctability of \cite{KSR} to qudits.
It turns out that this generalisation is straightforward and essentially is a reformulation of
the previous result. The main difficulty in the proof lies in dealing with Shannon entropies for $d$ elements
instead of the binary Shannon entropy.
\par As in the qubit case we focus on Taylor expansions of the Shannon entropy. The following two lemmata
will considerably simplify our approach.
\begin{Lemma}[Bounds for the Shannon entropy]\label{ShSchr}\hfill\\
  Let $\ksi = (\ksi_0, \dots, \ksi_{d-1}) \in \cW_d$ and set $x_n := \sum_{i = 1}^{d-1} \ksi_i
  = 1 - \ksi_0$. If we associate to $\ksi$ the distributions $\ksi_{\min} := (\ksi_0, x_n, 0, \dots, 0)$
  and $\ksi_{\max} := (\ksi_0, \frac{x_n}{d-1}, \dots, \frac{x_n}{d-1})$, then
  \begin{equation*}
    H_d(\ksi_{\min}) \leq H_d(\ksi) \leq H_d(\ksi_{\max})
  \end{equation*}
  holds, and we calculate
  \begin{eqnarray*}
    H_d(\ksi_{\min}) &= -(\ln d)^{-1}\left[\ksi_0 \ln \ksi_0 + x_n \ln x_n\right],\\
    H_d(\ksi_{\max}) &= -(\ln d)^{-1}\left[\ksi_0 \ln \ksi_0 + x_n \ln \frac{x_n}{d-1}\right].
  \end{eqnarray*}
\end{Lemma}
\Proof See \ref{BewShSchr}.
\begin{Lemma}[A Taylor expansion for the Shannon entropy]\label{Taylor}\hfill\\
  Let $p = (p_0, \dots, p_{d-1}) \in \cW_d$ and denote by $g := (1/d, \dots, 1/d) \in \cW_d$
  the uniform probability distribution on a set with $d$ elements. Provided that there
  exists some factor $f > 0$, such that $p_i \geq f/d$ holds for all $i$, we have
  \begin{equation*}
    H_d(p) = 1 - K\norm{g-p}_2^2 + K^\prime\eps(p)\norm{g-p}_3^3
  \end{equation*}
  for some $K,\,K^\prime > 0$ and a bounded function \Fkt{\eps}{\cW_d}{[-1;1]}.
\end{Lemma}
\Proof See \ref{BewTaylor}.\vspace{0.5cm}\\
The following theorem now generalises Theorem 1 of \cite{KSR} to higher dimensions.
\begin{Theorem}[Asymptotic correctability]\label{AsymCorr}\hfill\\
  Let $d$ be prime and $\rho = (A_{lm})_{l,m=0}^{d-1} \in \Sbdd$ be a state
  on which for each $n \in \N$ a (fictive) $\Snd$ step is applied to; the resulting state
  shall be called $(A^\prime_{lm})_{l,m=0}^{d-1} \in \Sbdd$. Define by
  \begin{itemize}
    \item $x_n := \sum_{m=1}^{d-1} \sum_{l=0}^{d-1} A^\prime_{lm}$ the total dit-error rate;
    \item $y_n := \norm{g - p}_2 / \sqrt{2}$ a measure for the deviation of the phase error probability
      $p = (A^\prime_{l*})_{l = 0}^{d-1}$ from the uniform probability distribution
      $g = (1/d, \dots, 1/d)$.\footnote{The factor $\sqrt{2}$ next to $y_n$ is only for consistency
      of notation with the qubit case \cite{KSR}.}
  \end{itemize}
  Provided that the sequence \Folge{x}{n}{\N}\, converges to zero, we have
  \begin{enumerate}
    \item If there exists an $r > 2$ such that $\sup\Mge{x_n/y_n^r}{n \in \N} < \infty$, then
      $\rho$ is asymptotically $S_n$-correctable.
    \item If, on the other hand $\inf\Mge{x_n/y_n^2}{n \in \N} > 0$ holds, then $\rho$ is asymptotically
      non-correctable with respect to that sequence.
  \end{enumerate}
  Both statements remain valid, if the role of dit errors and phase errors is interchanged.
\end{Theorem}
\Proof We may assume that $\lim_\nGr y_n = 0$; otherwise our statement follows directly from Theorem \ref{QSS}.
Considering the distribution of dit errors $\ksi = (A_{*m})_{m = 0}^{d-1}$ and using the binary Shannon
entropy $H(x) = - x \log_2 x - (1-x) \log_2 (1-x)$, Lemma \ref{ShSchr} allows us to write
\begin{equation}
  H_d(\ksi) = L \cdot H(x_n) + c(\ksi)\,x_n,
\end{equation}
where $L = \ln 2 / \ln d$ and $\Fkt{c}{\cW}{[0;\log_d (d-1)]} \subseteq [0;1]$ is some bounded function.
By Lemma \ref{Taylor}, for the distribution of phase errors $p$ due to
$(2y_n^2)^{3/2} = \norm{g-p}_2^3 \geq \norm{g-p}_3^3$ we have
\begin{eqnarray}
  H_d(p) &= 1 - K \cdot 2y_n^2 + K^\prime \eps(p) \norm{g-p}_3^3 \\
         &= 1 - K \cdot 2y_n^2 + K^\prime \eps^\prime(p) \cdot (2y_n^2)^{3/2},
\end{eqnarray}
where $K,\,K^\prime > 0$ and \Fkt{\eps,\,\eps^\prime}{\cW_d}{[-1;1]}\, are bounded functions, provided
$y_n$ is sufficiently small. Setting $\rho^\prime := \Aplm$ yields
\begin{eqnarray}
  \AsymCSS(\rho^\prime) &= 1 - H_d(\ksi) - H_d(p) \\
    &= - L \cdot H(x_n) - c(\ksi)\,x_n + 2K \cdot y_n^2
       - 2\sqrt{2}K^\prime \eps^\prime(p) \cdot y_n^3,
\end{eqnarray}
that is
\begin{equation}\label{AKhinr}
  \AsymCSS(\rho^\prime) > 0 \eqv \frac{-L \cdot H(x_n)}{y_n^2}
    - c(\ksi)\,\frac{x_n}{y_n^2} + 2K - 2\sqrt{2}K^\prime \eps^\prime(p) \cdot y_n > 0.
\end{equation}
In the following, we will also use the property that $\lim_{x \rightarrow 0^+} H(x)/x^s = 0$ for
$s \in [0;1)$ and $\lim_{x \rightarrow 0^+} H(x)/x^s = +\infty$ for $s \in [1;\infty)$.
\par For the proof of statement (i), note that condition (i) now implies that $x_n \leq c y_n^r$
for some $c \geq 0$, which yields $-L \cdot H(x_n)/y_n^2 \leq -L \cdot c^{2/r} H(x_n)/x^{2/r} \rightarrow 0$
for \nGr\, due to $r > 2$. This means that in (\ref{AKhinr}) all terms except $2K$ converge to zero.
For the proof of (ii), we have $x_n \geq c y_n^2$ for some $c$. In a similar fashion as before, this
results in $-L \cdot H(x_n)/y_n^2 \geq -L \cdot c H(x_n)/x \rightarrow -\infty$. Also, the second term
is negative, whereas all other terms are bounded, so that for sufficiently large $n$ the quantum Shannon
bound is not fulfilled. \BE

\section{Entanglement purification protocols and asymptotic correctability}\label{SecPurProt}
In this section, we want to apply our criterion to an actual sequence of correction steps. We therefore
focus on a well-known example for two-way entanglement purification, which we will call $\Bnd$ steps and
which are defined for any $n \in \N$. Considering a state $\rho = \Alm \in \Sbd$, the main objective
of this section is to derive a condition on $\rho$ for asymptotic \Bnd-correctability. It will turn out,
that we can calculate a \emph{characteristic exponent} $r^{(d)}$, such that for the case $r^{(d)} > 2$
we have asymptotic \Bnd-correctability, whereas for $r^{(d)} \leq 2$ we have non-correctability. These
results generalise our previous results from \cite{KSR} from qubits to qudits.

\subsection{Bell diagonal states and \Bnd steps}
We now introduce a generalisation of the \Bn\, step to $d$ dimensions. For $n \in \N$, a \Bnd\, step
is defined by the following procedure.
\begin{enumerate}
  \item Alice and Bob arbitrarily choose $n$ qudit pairs $QP_1, \dots, QP_n$.
  \item Alice and Bob apply $n-1$ $\GBXOR$ transformations with control $QP_1$ and target pairs
    $QP_2, \dots, QP_n$.
  \item Alice and Bob measure the parity on the pairs $QP_2, \dots, QP_n$ and discard the measured pairs.
    They keep $QP_1$, if and only if all parities are zero, otherwise they discard it.
\end{enumerate}
Starting with a tensor product of Bell states, the transformation of step (ii) is given by
\begin{equation}
  \bigotimes_{i = 1}^{n}\,\,\lm{l_i}{m_i} \mapsto \lm{\bigoplus_{i = 1}^{n} l_i}{m_1}
    \x \left[\bigotimes_{k = 2}^{n} \lm{l_k}{m_1 \ominus m_k}\right].
\end{equation}
The first pair is thus kept, if $m_1 \ominus m_k = 0$ holds for all $k \in \Mg{2,\dots,n}$.
\par Because we deal with mixtures of generalised Bell states, we want to formulate a $\Bnd$ step as
a mapping on the set \Sbdd. This is done in the following theorem.
\begin{Theorem}[Evolution of states for \Bnd\, steps]\label{BndEntw}\hfill\\
  For each $k \in \MgE{n}$ let $\rho^{(k)} = (A_{lm}^{(k)})_{l,m = 0}^{d-1} \in \Sbdd$ be a state.
  If a \Bnd\, step is applied to these states and if not all pairs are discarded, the
  state of the remaining pair is given by $\rho^\prime = \Aplm \in \Sbdd$ with coefficients
  \begin{equation*}
    A_{lm}^\prime = (dN)^{-1} \sum_{i = 0}^{d-1} \, \Biggl[ \, z^{-il} \,
      \prod_{k = 1}^{n} \,\, \biggl(\sum_{j = 0}^{d-1} z^{ij} A_{jm}^{(k)} \biggr)\Biggr],
  \end{equation*}
  where $z := \exp(2\pi\iE/d)$\, denotes the principal value of the root of unity of order $d$ and
  $N := \sum_{m = 0}^{d-1} \bigl[\prod_{k = 1}^n (\sum_{l = 0}^{d-1} A_{lm}^{(k)})\bigr]$ is
  the normalisation constant, i.\,e. the probability of survival of the first qudit pair.
  Note that the final state is itself Bell diagonal and does not depend on the ordering of the
  initial states.
\end{Theorem}
\Proof See \ref{BewBndEntw}.\\
Although we will not use it, it may be worth mentioning that a sequence of a $\Bnd$ step and a
$B_m^{(d)}$ step is equivalent to a single $B_{n \cdot m}^{(d)}$ step.

\subsection{Asymptotic correctability using a sequence of \Bnd\, steps}
Before we proceed with the calculation, we have to introduce some notation. As might be seen from
Theorem \ref{AsymCorr}, we mainly have to focus on purely exponential behaviour, that is, in many equations
we will skip subexponential terms. To be precise, for some non-negative-valued function, we define
its exponent by $z(f) := \lim_\nGr \sqrt[n]{f(n)}$, where we always assume that this limit exists;
any such function may now be written as $f(n) = c(n) z^n$ for some subexponential function $c$,
i.\,e. some function $c$ such that $z(c) = 1$ holds. We call two-functions $f$ and $g$
\emph{asymptotically exponentially equal}, if $z(f) = z(g)$, in which case we shall write $f \aeg g$.
\par For simplicity we will further assume that $A_{*0} > \max\Mge{A_{*m}}{m \in \Z_d^*}$ holds;
if this is not the case, we can apply the local-unitary operation $\Eins \x \sum_{k \in \Z_d}
\ket{k \ominus m}\bra{k}$, provided that $A_{*m}$
is the unique largest column. We further assume that the phase error rates converge to the uniform
probability distribution, which is always the case unless the the component of the Fourier transform
of the first column which has maximum absolute value is not unique.

\subsection{Evolution of dit errors}
The evolution of dit errors is straightforward. We denote by $\ksi = (\ksi_0, \dots, \ksi_{d-1}) \in \cW_d$
the distribution of dit errors, i.\,e. $\ksi_m := A_{*m}$ for $m \in \Z_d$. The application of a \Bnd\, step
may be viewed as a mapping $\Bnd : \ksi \mapsto \ksi^\prime$, defined by
\begin{equation}
  \ksi^\prime_i = \frac{\ksi_i^n}{\ksi_0^n + \dots + \ksi_{d-1}^n} \qquad\mathrm{for}\quad i \in \Z_d,
\end{equation}
which follows directly from Theorem \ref{BndEntw}. Therefore, using the notation of Theorem \ref{AsymCorr},
\begin{equation}
  x_n := 1 - \ksi^\prime_0 = \frac{\sum_{m = 1}^{d-1} \ksi_m^n}{\sum_{m = 0}^{d-1} \ksi_m^n}
    = \left[\frac{\sum_{m = 0}^{d-1} \ksi_m^n}{\sum_{m = 1}^{d-1} \ksi_m^n}\right]^{-1}
    = \left[1 + \frac{\ksi_0^n}{\sum_{m = 1}^{d-1} \ksi_m^n}\right]^{-1}.
\end{equation}
Setting $\ksi_{\max} := \max\Mge{\ksi_m}{m \in \Z_d^*}$, the following inequality holds for the
denominator:
\begin{equation}
  \ksi_{\max}^n \leq \sum_{m = 1}^{d-1} \ksi_m^n \leq (d-1) \ksi_{\max}^n.
\end{equation}
Using an appropriate function $\Fkt{h}{\cW_d}{[1;d-1]}$ yields
\begin{equation}\label{Def_un}
  x_n = \left[1 + \frac{\ksi_0^n}{h(\ksi)\ksi_{\max}^n}\right]^{-1} =: u(n) \tilde x^{-n},
\end{equation}
where $\tilde x := \ksi_0 / \ksi_{\max} > 1$ and appropriate values $u(n) \in [1/2;d]$. In particular,
we have $x_n \aeg \tilde x^{-n}$ and $\lim_\nGr \ksi_m^\prime = \delta_{m,0}$, so that the correction
of dit errors is guaranteed under $\Bnd$ steps.

\subsection{The evolution of phase errors}
In comparison to the dit-error evolution, the calculation of the phase errors is more sophisticated.
For using Theorem \ref{AsymCorr}, we only need to calculate the value $2y_n^2 = \norm{g-p}_2^2$,
where $p$ is the phase error distribution ($p_l := A_{l*}$) and $g = (1/d, \dots, 1/d)$ is the uniform
probability distribution. By use of Theorem \ref{BndEntw}, it follows
\begin{equation}
  2y_n^2 = \norm{g-p}_2^2
    = \norm{\left(\frac{1}{d} - \frac{\sum_m \sum_i z^{-il} \left(\sum_j z^{ij} A_{jm}\right)^n}{dN}
    \right)_{l = 0}^{d-1}}_2^2.
\end{equation}
The 2-norm is invariant with respect to a discrete Fourier transform
$(x_i)_i \mapsto (d^{-1/2}\sum_{i = 0}^{d-1} z^{ij} x_i)_j$. Thus the use of $\sum_{i = 0}^{d-1} z^{ik}
= d\delta_{k,0}$ implies
\begin{equation}
  2y_n^2 = \frac{1}{d} \norm{\left(\delta_{l,0} - \frac{\sum_m \left(\sum_j z^{lj} A_{jm}\right)^n}{N}
    \right)_{l = 0}^{d-1}}_2^2.
\end{equation}
The zero component cancels against the normalisation; this yields
\begin{equation}\label{ynErg}
  2y_n^2 = \frac{1}{d} \norm{\left(\frac{\sum_m \left(\sum_j z^{lj} A_{jm}\right)^n}{N}
    \right)_{l = 1}^{d-1}}_2^2,
\end{equation}
where we take the 2-norm on $d-1$ elements only.
The evaluation in the general case is complicated, although one may expect that in the limit $\nGr$
only the first column of \Alm\, should be relevant. In the next section we will slightly modify
the protocol, so that a calculation of the exponential behaviour of $2y_n^2$ for the modified protocol
becomes possible.

\subsection{The mixing operation}
Consider the single-qudit transformation $U_1 := d^{-1/2} \sum_{x = 0}^{d-1} z^{-x^2} \pr{x}$ and define
\begin{equation}
  U := U_1 \x U_1^* = \sum_{x,y = 0}^{d-1} z^{y^2- x^2} \pr{x} \x \pr{y}.
\end{equation}
This implies $U\ket{\Psi_{lm}} = z^{m^2} \ket{\Psi_{l\ominus 2m, m}}$ or $U : (l,m) \mapsto (l \ominus 2m,m)$.
That is, the transformation of a Bell-diagonal state by the local unitary operation $U$ permutes
the coefficients within a fixed column of the coefficient matrix. This
property can be used to simplify the calculation of $2y_n^2$; we therefore introduce the following
step immediately before Alice and Bob apply the \Bnd\, step.
\begin{itemize}
  \item For each qudit pair Alice and Bob randomly choose a value $n \in \Z_d$ and apply $U^n$
    to the respective pair.
\end{itemize}
Considering a density matrix $\rho$, this means $\rho \mapsto d^{-1} \sum_{n = 0}^{d-1}
U^n \rho (U^\dagger)^n$. For a mixture of Bell states, $\rho = \Alm$, this step mixes the entries
in the columns. Complete mixing within column $m$, i.\,e. $A_{lm} \mapsto A_{*m}/d$,
will take place, if $2m$ and the dimension $d$ are coprime. If we want to have complete mixing for all
columns except the $m = 0$ column, we have to restrict our considerations to odd primes (which we already
did due to the use of Theorem \ref{QSS}); the case $d = 2$ (the only even prime) was done in \cite{KSR}.
\par For fixed $l \in \Z_d$, one can calculate
\begin{eqnarray}
  \sum_{m = 0}^{d-1} \left(\sum_{j = 0}^{d-1} z^{lj} A_{jm}\right)^n
  &= \left(\sum_{j = 0}^{d-1} z^{lj} A_{j0}\right)^n
     + \sum_{m = 1}^{d-1} \left(\sum_{j = 0}^{d-1} z^{lj} \frac{A_{*m}}{d} \right)^n \\
  &= \left(\sum_{j = 0}^{d-1} z^{lj} A_{j0}\right)^n + \sum_{m = 1}^{d-1} \left(\frac{A_{*m}}{d}\right)^n
     \underbrace{\left(\sum_{j = 0}^{d-1} z^{lj} \right)^n}_{= d \cdot \delta_{l,0}}
\end{eqnarray}
and due to $l \neq 0$ in (\ref{ynErg}) it follows
\begin{equation}
  2y_n^2 \cdot dN^2 = \norm{\left(\left(\sum\nolimits_j z^{lj} A_{j0}\right)^n \right)_{l = 1}^{d-1}}_2^2
                    = \norm{\left(\sum\nolimits_j z^{lj} A_{j0} \right)_{l = 1}^{d-1}}_{2n}^{2n}.
\end{equation}
It can now be seen, that $\norm{x}_{2n}^{2n} = K(n) \norm{x}_\infty^n$ for any $d$-tuple $x$,
where $K(n) \in [1;d]$ may depend on $x$. This yields
\begin{eqnarray}
  2y_n^2 \cdot dN^2 &= K(n) \norm{\left(\sum\nolimits_j z^{lj} A_{j0} \right)_{l = 1}^{d-1}}_{\infty}^{2n}\\
    &= K(n) \left[\max\Mge{\betrag{\sum\nolimits_j z^{lj} A_{j0}}}{l \in \Z_d^*}\right]^{2n}.
\end{eqnarray}
This shows that the determination of the evolution of phase errors is related to the search for the
largest absolute value of the Fourier transform of a probability distribution, where the zero component
of the transformed tuple is ignored.

\subsection{Exponential behaviour}
Up to now, we have shown  $x_n \aeg \tilde x^{-n}$, where $\tilde x = A_{*0}/\max \Mge{A_{*m}}{m \in \Z_d^*}$.
This implies for the normalisation constant of a $\Bnd$ step that $N_n = K^\prime(n)\,A_{*0}^n$ for
$K^\prime(n) \in [1;d]$. Thus we find $N_n \aeg A_{*0}^n$ which implies
\begin{equation}\label{Def_yn}
  2y_n^2 \aeg \frac{1}{d} \cdot \frac{K(n)}{K^\prime(n)} \cdot
    \left(\frac{\max{\Mge{\betrag{\sum_j z^{lj} A_{j0}}}{l \in \Z_d^*}}}{A_{*0}}\right)^{2n}
    =: \frac{K(n)}{K^\prime(n)} \cdot \frac{\tilde y^{2n}}{d}.
\end{equation}
The condition $x_n \aeg y_n^{r^{(d)}}$ now yields $\tilde x^{-n} = \tilde y^{r^{(d)}n}$ or
\begin{equation}\label{rExp}
  r^{(d)} = - \frac{\ln \tilde x}{\ln \tilde y}
          = \frac{\ln\,\bigl[A_{*0} / \max \Mge{A_{*m}}{m \in \Z_d^*}\bigr]}
            {\ln\,\Bigl[A_{*0} / \max{\Mge{\betrag{\sum_j z^{lj} A_{j0}}}{l \in \Z_d^*}}\Bigr]}.
\end{equation}
This generalises the characteristic exponent $r$ from our previous work \cite{KSR} from qubits to
qudits.
\par Finally, we have to relate the characteristic exponent $r^{(d)}$ to the conditions in
Theorem~\ref{AsymCorr}; this we will do in the following theorem.
\begin{Theorem}[Asymptotical \Bnd-correctability]\label{BndCorr}\hfill\\
  A state $\rho = \Alm \in \Sbdd$ is asymptotically \Bnd-correctable, if and only if $r^{(d)} > 2$.
\end{Theorem}
\Proof Setting $r := r^{(d)}$ and using (\ref{Def_un}) and (\ref{Def_yn}) we find
\begin{equation}
  \frac{x_n}{y_n^r} = u(n) \tilde x^{-1} \cdot \left(\tilde y^{2n}
    \frac{K(n)}{2dK^\prime(n)} \right)^{-r/2}
    = \frac{u(n)}{(\tilde x \cdot \tilde y^r)^n} \left(\frac{K(n)}{2dK^\prime(n)} \right)^{-r/2}.
\end{equation}
The characteristic exponent $r^{(d)}$ is chosen in such a way that $(\tilde x \cdot \tilde y^r)^n = 1$
(in particular, $x_n/y_n^r \aeg 1$). The remaining terms are bounded for all $n \in \N$ by some
lower bound being larger than zero and some upper bound being less than infinity. Thus,
Theorem~\ref{AsymCorr} implies the assertion. \BE

\section{Applications in quantum cryptography}\label{SecQKD}
Let us now consider some cryptographical applications of our theorems. In the generic model of
entanglement-based quantum cryptography, Alice prepares the state $\ket{\Psi_{00}}^{\x n}$ and sends every
second qudit to Bob. The transmission is considered to be insecure, so that Eve can perform general coherent
attacks.
The task of Alice and Bob is now to estimate the resulting errors and, if possible, to perform entanglement
purification. This provides Alice and Bob with (nearly) maximally entangled states, from which they can extract
a secret key.
\par Although in general the total state of Alice and Bob is complicated, a random permutation of their
qudit pairs and a fictive-Bell-measurement argument \cite{GL} allows us to restrict the theoretical analysis
to tensor products of mixtures of generalised Bell states. If we consider protocols consisting of one
$\Bnd$ step for an appropriately chosen $n \in \N$ and the application of a CSS code according
to Theorem~\ref{QSS}, we only have to determine the coefficients $\Alm$ in order to determine,
whether we can obtain a secret key.
\par A final remark has to be made on prepare-and-measure protocols. The reduction of CSS-based protocols
for qudits was done by Hamada \cite{Ham} and the reduction of $\Bnd$ steps also follows the well-known
lines (cf. e.\,g. \cite{GL,NA}). The only remaining point is the reduction of our mixing operation; but
this mixing only mixes phases and does not change any dit value and therefore has no influence on the key.
This means Alice and Bob can just skip it in the associated prepare-and-measure protocol.
\par In the remaining part we will consider states which may appear in a quantum cryptographic protocol,
and we will also deal with the problem that in general we cannot infer all coefficients $\Alm$ from
measurements.

\subsection{The generalised isotropic case}
We start with a particularly simple example, namely \emph{generalised isotropic states}, which were also
considered in \cite{NRA}. A generalised isotropic state is of the form
\begin{equation}\label{isotrop}
  \rho = (\alpha,\beta,\gamma,\delta) := \left(\begin{array}{cccc}
    \alpha & \gamma & \dots  & \gamma \\
    \beta  & \delta & \dots  & \delta \\
    \vdots & \vdots & \ddots & \vdots \\
    \beta  & \delta & \dots  & \delta \\
  \end{array}\right) \in \Sbdd.
\end{equation}
If $\beta = \gamma$, this is called an \emph{isotropic state}. An interesting property of generalised
isotropic states is that they remain of this form, if they are subjected to \Bnd\, steps; it is thus
possible to view a $\Bnd$ step as a mapping $\Bnd : (\alpha,\beta,\gamma,\delta) \mapsto
(\alpha^\prime, \beta^\prime, \gamma^\prime, \delta^\prime)$, where the coefficients are given by
\begin{equation}\begin{array}{rlr}
  \alpha^\prime &= \left\{\bigl[\alpha+(d-1)\beta\bigr]^n  + (d-1)\bigl[\alpha-\beta\bigr]^n\right\} / dN,\\
  \beta^\prime  &= \left\{\bigl[\alpha+(d-1)\beta\bigr]^n  - \bigl[\alpha-\beta\bigr]^n\right\} / dN,\\
  \gamma^\prime &= \left\{\bigl[\gamma+(d-1)\delta\bigr]^n + (d-1)\bigl[\gamma-\delta\bigr]^n\right\} / dN,\\
  \delta^\prime &= \left\{\bigl[\gamma+(d-1)\delta\bigr]^n - \bigl[\gamma-\delta\bigr]^n\right\} / dN,\\
  N             &= \bigl[\alpha+(d-1)\beta\bigr]^n         + (d-1)\bigl[\gamma+(d-1)\delta\bigr]^n.
\end{array}\end{equation}
Evaluation of (\ref{rExp}) now yields
\begin{equation}
  r^{(d)} = \left[\ln \frac{\alpha+(d-1)\beta}{\gamma+(d-1)\delta}\right]
            \big/ \ln \left[\frac{\alpha+(d-1)\beta}{\betrag{\alpha-\beta}}\right],
\end{equation}
and thus $r^{(d)} > 2 \eqv \alpha^2 + \beta^2 - 2[\alpha + (d-1)\beta]/d > 0$. Using $\alpha > \beta$,
we regain the result for isotropic channels of our previous work \cite{NRA}.
In the case $d = 2$, this state reduces to the general mixture of qubit Bell states as considered
in \cite{KSR}. Further note, that in the case of generalised isotropic channels we could have done
the calculation for $r^{(d)}$ without the use of the mixing operation.

\subsection{Maximum tolerable error rates for two-basis cryptography}
In quantum cryptography, the protocol in \cite{Ch05} produces isotropic states, where
$\beta = \gamma = \delta$, but uses $d+1$ mutually orthogonal bases. On the other hand, the theoretical
analysis of protocols which use only two bases do not, in general, leads to generalised isotropic states.
\par Let us now focus on protocols which use two Fourier-dual bases and in which the the total dit value
probabilities $A_{*m}$ ($m \in \Z_d$) are measured. Such protocols were considered
in \cite{NA} and it was shown there, that for $l,\,m \in \Z_d$ the symmetry relations
\begin{equation}
  A_{lm} = A_{d-m,l} = A_{d-l,d-m} = A_{m,d-l}
\end{equation}
hold for the quantum states describing Alice's and Bob's entanglement. A consequence of these relations
is $A_{l*} = A_{*l}$ for $l \in \Z_d$.
\par From the measured dit errors $A_{*m}$ a lower bound on $r^{(d)}$ has to be inferred.
From~(\ref{rExp}) it can be seen that we need three quantities to calculate $r^{(d)}$, namely $x := A_{*m}$,
$\max\Mge{A_{*m}}{m \in \Z_d^*}$ and
\begin{equation}
  M := \max{\Mge{\betrag{\sum\nolimits_{j = 0}^{d-1} z^{lj} A_{j0}}}{l \in \Z_d^*}}.
\end{equation}
We will write $\max\Mge{A_{*m}}{m \in \Z_d^*} = f \cdot (1-x) \cdot (d-1)^{-1}$, where $f \in [1;d-1]$.
The case $f = 1$ is the \emph{apparently isotropic} case, where all $A_{*m}$ for $m \in \Z_d^*$ are equal,
whereas $f = d-1$ relates to those cases, in which there are only errors of one type.
Equation~(\ref{rExp}) now reads
\begin{equation}
  r^{(d)} = \left(\ln \frac{x}{f \cdot \frac{1-x}{d-1}}\right) \cdot \left(\ln \frac{x}{M}\right)^{-1}.
\end{equation}
The values of $x$ and $f$ can be directly inferred from the measured dit-error probabilities. However,
estimating the value of $M$ is more involved.
We note that small values of $M$ correspond to small values of $r^{(d)}$. So, for a lower bound on $r^{(d)}$
we need a lower bound on $M$, which will be derived now.
\par For any complex number $z \in \C$, we have $\betrag{z} \geq \mathrm{Re}\,z$ and the maximum over all
$l \in \Z_d^*$ is definitely larger than the average over this set. We thus have
\begin{equation}
  M \geq \max{\Mge{\mathrm{Re}\sum\nolimits_{j = 0}^{d-1} z^{lj} A_{j0}}{l \in \Z_d^*}}
    \geq \mathrm{Re} \frac{1}{d-1}\sum\nolimits_{l = 1}^{d-1} \sum\nolimits_{j = 0}^{d-1} z^{lj} A_{j0}.
\end{equation}
Exchanging the summation and using the fact that $\sum\nolimits_{l = 1}^{d-1} z^{lj} = d\delta_{j,0} - 1$
yields
\begin{equation}\label{M1}
  M \geq \frac{1}{d-1} \sum\nolimits_{j = 0}^{d-1} (d\delta_{j0} - 1) A_{j0}
    = A_{00} - \frac{\sum_{j = 1}^{d-1} A_{j0}}{d-1}.
\end{equation}
Note that in the case of the generalised isotropic channel this is an equality. Up to this point
we have given a simple, but achievable lower bound on $M$. In order to infer this lower bound from the
qudit-error probabilities measurable in the protocol we use the relations
\begin{equation}\label{M2}
  A_{*0} = A_{00} + \sum_{l = 1}^{d-1} A_{l0} \leq A_{00} + \sum_{l = 1}^{d-1} A_{l*}
         = A_{00} + \sum_{l = 1}^{d-1} A_{*l} = A_{00} + (1 - A_{*0}),
\end{equation}
which imply $A_{00} \geq 2 A_{*0} - 1$. Note that equality holds, if and only if $A_{lm} = 0$ for
$(l,m) \in \Z_d^* \times Z_d^*$. Plugging this bound into the bound for $M$ yields
\begin{equation}\label{M3}
  M \geq x - d \cdot \frac{1-x}{d-1}.
\end{equation}
The isotropic channel of (\ref{isotrop}) is the worst case with respect to correctability (i.\,e., it has
the smallest $r^{(d)}$) of all apparently isotropic channels, i.\,e. channels where $A_{*m} = A_{*m^\prime}$
for all $m,\,m^\prime \in Z_d^*$.
Furthermore, we have equality in (\ref{M2}) and thus in (\ref{M3}), if for this
isotropic channel $\delta = 0$ holds; this case was considered in \cite{NRA}.
If we do not have an isotropic channel, the tolerable error rate according to our bound depends
on $f$, which can be seen as a parameter characterizing the non-isotropy of the measured probability
distribution.
\par
\begin{figure}[t]\begin{center}
  \includegraphics[height=145pt,width=250pt]{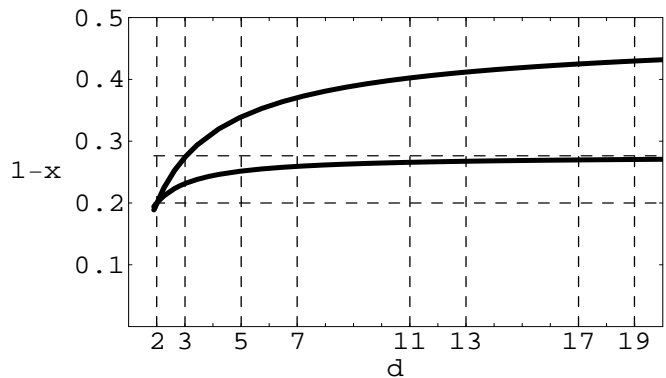}
  \caption{Lower bounds for the maximum tolerable error rate ($1-x = 1-A_{*0}$) as a function of the
    dimension $d$; the upper line corresponds to the apparently isotropic case $f = 1$ (where this
    bound is exact), the lower one to the maximum non-isotropy $f = d-1$. All other cases lie in between.
    The lines start at $1 - x = 0.2$, the upper one converges to $0.5$, the lower one to
    $1/2- 1/2\sqrt{5} \approx 0.276$.}
  \label{Bounds}
\end{center}\end{figure}

 By plugging in our bound for $M$ and solving for $x = A_{*0}$, we get as a sufficient condition for
correctability
\begin{equation}
  x > \frac{2d(2d-1) + (d-1)(f + \sqrt{(4d+f)f})}{2[(2d-1)^2 + (d-1)f]},
\end{equation}
where we only consider $x > (d+1)/(2d)$ due to the entanglement bound of \cite{NA}.
In figure~\ref{Bounds} we plotted bounds on the maximum tolerable error rate ($1-x$) as a function of $d$.
The upper line is the apparently isotropic case ($f = 1$), the lower one the case with just one type
of error ($f = d-1$). The lower bound for the maximum tolerable error rate in a given protocol lies
between these two lines.
We thus have shown lower bounds on the maximum tolerable error rates of two-basis quantum cryptography
using the protocols considered. In case of apparently isotropic channels our bounds are exact lower
bounds, in other cases they become worse the more non-isotropic the channel gets.

\section{Conclusions}
We have generalised the ideas of our previous work \cite{KSR}, namely the notion of asymptotic
correctability, to $d$-dimensional quantum systems. We determined a criterion for asymptotic correctability
and applied it to $\Bnd$ steps, which yielded an expression for the characteristic exponent $r^{(d)}$
related to asymptotic \Bnd-correctability. Applying this condition to cryptographic protocols yielded
lower bounds for maximum tolerable error rates and the bound in the case of apparently isotropic
channels is tight.
\par Up to now our proof relies on the existence of asymmetric CSS codes for $d$ dimensions~\cite{Ham}.
If such codes exist for non-prime dimensions (e.\,g. prime powers), our result can be easily generalised
to these dimensions, provided the mixing operation is adapted accordingly. It would also be interesting
to explicitly calculate the value $M = \max{\Mge{\betrag{\sum\nolimits_{j = 0}^{d-1} z^{lj}
A_{j0}}}{l \in \Z_d^*}}$, if $A_{l0}$ are known for all $l \in \Z_d$ and to infer better bounds on $M$ for
the non-isotropic case by using the symmetry relations of two-bases protocols, but both tasks seem to be
relatively complicated.

\ack This work is supported by the EU within the IP SECOQC. Informative discussions with 
Georgios M. Nikolopoulos are acknowledged. K. S. Ranade is supported by a graduate-student
scholarship of the Technische Universit\"at Darmstadt.

\appendix
\section{Proof of Lemma \ref{ShSchr}}\label{BewShSchr}
By definition of the Shannon entropy, it is obvious that it is invariant with respect to any
permutation of the $\ksi_i$. Furthermore, we know that it is concave, i.\,e.
\begin{equation}
  H_d(\lambda \cdot \ksi + (1-\lambda) \cdot \eta) \geq \lambda H_d(\ksi) + (1-\lambda) H_d(\eta)
  \qquad\mathrm{for}\quad \lambda \in [0;1].
\end{equation}
One now can see that $\ksi_{\max}$ can be represented as a mixture of permutations of $\ksi$, where
$\ksi_0$ is left invariant, and, on the other hand, $\ksi$ can be constructed by a mixture of
permutations of $\ksi_{\min}$. \BE

\section{Proof of Lemma \ref{Taylor}}\label{BewTaylor}
A Taylor expansion of $H_d$ up to second order around $g$ yields
\begin{equation}
  H_d(p) = 1 + \sum_{i = 0}^{d-1} \left(1 - \frac{1}{\ln d}\right)(p_i - g_i)
    - \frac{d}{2 \ln d} \sum_{i = 0}^{d-1} (g_i-p_i)^2 + R_2(p).
\end{equation}
Due to the fact that we only consider probability distributions $p$, the first order term vanishes
and the second order term can be written in the form of Lemma \ref{Taylor} using $K := d/(2 \ln d)$.
The remainder term $R_2(p)$ can be calculated by Lagrange's formula, i.\,e.
\begin{equation}
  R_2(p) = \sum_{i = 0}^{d-1} \frac{\tilde p_i^{-2}}{3! \cdot \ln d} (p_i - g_i)^3
\end{equation}
for some set $\tilde p_i$, where $p_i \leq \tilde p_i \leq 1/d$ or $1/d \leq \tilde p_i \leq p_i$
holds for any $i$. By assumption, we have $\tilde p_i \geq f/d$; this yields
\begin{equation}
  \betrag{R_2(p)} \leq \sum_{i = 0}^{d-1} \frac{(f/d)^{-2}}{3! \cdot \ln d} (p_i - g_i)^3
    \leq K^\prime \norm{p-g}_3^3
\end{equation}
for $K^\prime := d^2 \cdot (3! f^2 \cdot \ln d)^{-1}$, which concludes the proof. \BE

\section{Proof of Theorem \ref{BndEntw}}\label{BewBndEntw}
In this section, we give the proof of Theorem \ref{BndEntw}, which closely follows the ideas
presented in \cite{MDN}. The main idea in the proof is that the phase propagation can be seen
as a convolution, which can be calculated by a sequence of Fourier transform, multiplication
and inverse Fourier transform.
\par The proof is done by induction, which (the case $n = 1$ being obvious) we start for $n = 2$.
Consider $(A_{lm})_{lm}, (B_{st})_{st} \in \Sbdd$ and denote $(l,m) := \pr{\Psi_{lm}}$. Applying
steps (i) and (ii) of a \Bnd\, step in this case yields
\begin{eqnarray}
  \rho &= \sum_{l,m} A_{lm} (l,m) \x \sum_{s,t} B_{st} (s,t)
       = \sum_{l,m,s,t} A_{lm} B_{st} (l,m) \x (s,t)\\
       &\stackrel{\GBXOR}{\mapsto} \sum_{l,m,s,t} A_{lm} B_{st} (l \oplus s, m) \x (s, m \ominus t).
\end{eqnarray}
Considering only the case where $m \ominus t = 0$ and tracing out the second pair further yields
\begin{eqnarray}
  N_2^{-1} \sum_{l,m,s} A_{lm} B_{sm} (l \oplus s, m)
    = \sum_{lm} \left[N_2^{-1} \sum_{l^\prime} A_{lm} B_{l \ominus l^\prime, m}\right] (l,m),
\end{eqnarray}
where $N_2 = \sum_{m}\left[(\sum_l A_{lm})(\sum_l B_{lm})\right]$ is the normalisation constant.
We assume now that the theorem is true for all numbers upto a fixed value $n$ and proceed via
induction: Let $\rho^{(i)} = (A_{lm}^{(i)})_{l,m=0}^{d-1}$ be mixtures of Bell states for
$i \in \MgE{n+1}$. The outcome of a \Bnd\, step applied to the states $1, \dots, n$ shall be denoted
as $\rho^\prime = \Aplm$ with normalisation constant $N_n$, the outcome of a $B_{n+1}^{(d)}$ on
all $n+1$ states shall be $\rho^{\prime\prime} = (A_{lm}^{\prime\prime})_{l,m=0}^{d-1}$. We calculate
\begin{eqnarray}
  A_{lm}^{\prime\prime}
    &= \frac{1}{d N_2} \sum_i z^{-il} \left[
         \left(\sum_j z^{ij} A_{jm}^\prime\right)
         \left(\sum_{j^\prime} z^{ij^\prime} A_{j^\prime m}^{(n+1)}\right)
       \right]\\
    &= \frac{1}{d^2 N_2 N_n} \sum_i z^{-il} \left[
         \sum_{i^\prime, j} z^{ij + i^\prime j} \prod_{k = 1}^{n+1}
         \left(\sum_{j^{\prime}} z^{i^\prime j^{\prime}} A_{j^{\prime}m}^{(k)}\right)
       \right]\\
    &= \frac{1}{d^2 N_2 N_n} \sum_{i, i^\prime, j} z^{i(j-l) + i^\prime j}
         \prod_{k = 1}^{n+1} \left(\sum_{j^{\prime}} z^{i^\prime j^{\prime}} A_{j^{\prime}m}^{(k)}\right),
\end{eqnarray}
where $N_2$ is the normalisation constant for a \Bnd\, step with $n = 2$ applied to $\rho^\prime$
and $\rho^{(n+1)}$. Using $\sum_{i = 0}^{d-1} z^{i(j-l)} = d \delta_{j,l}$, this implies the assertion,
if the normalisation constant is correct. This can be verified by direct calculation. \BE

\end{document}